\documentclass[prl,aps,superscriptaddress,amsmath,twocolumn]{revtex4-1}
\usepackage{graphicx}
\usepackage{dcolumn}
\usepackage{bm}
\usepackage{color}

\begin{document}

\title{Dissociation of high-pressure solid molecular hydrogen: 
       Quantum Monte Carlo and anharmonic vibrational study}

\author{Sam Azadi}%
\email{s.azadi@imperial.ac.uk}
\affiliation{Thomas Young Centre and Department of Physics, Imperial
  College London, London SW7 2AZ, United Kingdom}
\author{Bartomeu Monserrat}
\email{bm418@cam.ac.uk}
\affiliation{TCM Group, Cavendish Laboratory, University of Cambridge,
  J. J. Thomson Avenue, Cambridge CB3 0HE, United Kingdom}
\author{W. M. C. Foulkes}
\affiliation{Thomas Young Centre and Department of Physics, Imperial
  College London, London SW7 2AZ, United Kingdom}
\author{R. J. Needs}
\affiliation{TCM Group, Cavendish Laboratory, University of Cambridge,
  J. J. Thomson Avenue, Cambridge CB3 0HE, United Kingdom}
  
\date{\today}

\begin{abstract}

  A theoretical study is reported of the molecular-to-atomic transition
  in solid hydrogen at high pressure.  We use the diffusion quantum
  Monte Carlo method to calculate the static lattice energies of the
  competing phases and a density-functional-theory-based vibrational
  self-consistent field method to calculate anharmonic vibrational
  properties. We find a small but significant contribution to the
  vibrational energy from anharmonicity.  A transition from the
  molecular $Cmca$-$12$ direct to the atomic $I4_1/amd$ phase is found
  at $374$ GPa. The vibrational contribution lowers the transition
  pressure by $91$~GPa. The dissociation pressure is not very sensitive
  to the isotopic composition. Our results suggest that quantum melting
  occurs at finite temperature.

\end{abstract}

\maketitle


In 1935, Wigner and Huntington \cite{Wigner} predicted that solid molecular
hydrogen would dissociate at high pressure to form a metallic atomic solid. 
The properties of atomic hydrogen have fascinated
high-pressure scientists and astrophysicists ever since
\cite{McMahon_review_RMP,Goncharov_2013_review}. Various exotic
predictions have been made, such as the stability of atomic metallic
hydrogen in a superfluid state or as a room-temperature superconductor
\cite{Ashcroft1, Ashcroft1b, Ashcroft2}, but neither an
insulator-to-metal transition nor a molecular-to-atomic transition has
yet been observed unambiguously at low temperatures.

The nature of hydrogen at high pressure is currently the subject of
intense interest.  Experimental studies of hydrogen and deuterium have
been performed up to pressures above 300 GPa using static
diamond-anvil-cell (DAC) techniques
\cite{Eremets,Howie,Howie_2012,Hemley12,Loubeyre_2013,Goncharov_2013_review,Zha_2013}.
A new high-pressure phase IV of hydrogen and deuterium was recently
observed, which is believed to consist of alternate layers of strongly
bonded molecules and weakly bonded graphene-like sheets
\cite{Howie,Pickard_phase_IV}.  The precise pressures achieved in these
experiments may have been overestimated and are still controversial
\cite{Howie_pressure_measurement}. Even more controversial is the
suggestion that conductive dense hydrogen has been produced in
room-temperature experiments \cite{Eremets}. However, the discovery of
weak bonding in phase IV suggests that static DAC experiments could
probe the conditions under which full molecular dissociation of hydrogen
and deuterium take place. The results of our work corroborate this
suggestion.

We have studied hydrogen in the pressure range $300$--$650$ GPa,
within which the transition from molecular to atomic structures is
thought likely to occur.  The most important contribution to the
structural energy is the static lattice energy.  The energy
differences between competing phases in hydrogen are small and a very
accurate description of the electronic energy is required to resolve
them.  We have therefore calculated
static lattice energies using the diffusion quantum Monte Carlo (DMC)
method, which is the most accurate method known for evaluating the
energies of large assemblies of interacting quantum particles
\cite{Ceperley_DMC,matthew,Attacalite,Natoli93}.

Experimental measurements
\cite{Deemyad_2008,Eremets_melting_2009,Subramanian_melting_2011} and
classical molecular dynamics (MD) simulations using density functional
theory (DFT) methods \cite{Bonev_2004_MD,Tamblyn_2010_MD} suggest that
the melting temperature of hydrogen increases with pressure and
reaches a maximum value of roughly 1000 K at a pressure in the region
of 100 GPa, whereafter it declines with pressure.  Path-integral molecular dynamics (PIMD) simulations
have suggested that the inclusion of the zero-point (ZP) energy of the protons
reduces the melting temperature to about 160 K at 500 GPa and 100 K at
800 GPa \cite{Chen}.  This suggests that the inter-atomic bonding
becomes very weak at these pressures, and anharmonic effects could
become important.

We have performed vibrational self-consistent field (VSCF) calculations
within DFT to calculate the anharmonic vibrational ZP energies
\cite{Tomeu1}. 
We used the Perdew-Burke-Ernzerhof
(PBE) generalized gradient approximation density functional, which is
well suited for very high-pressure studies, as the charge density is
more uniform than at low densities, and it obeys the uniform limit and
gives a good account of the linear response of the electron gas to an
external potential \cite{PBE}.

Static lattice DFT calculations using {\it ab initio} random structure
searching (AIRSS) \cite{PickardReview} indicate that there are three
energetically competitive structures in the range of interest.
The molecular $Cmca$-$12$ phase is insulating up to
$373$~GPa in the $GW$ approximation~\cite{me3}, although proton
zero-point and finite-temperature effects are expected to lower the
metallization pressure~\cite{Morales_2013}. The molecular $Cmca$-4
phase and the atomic phase of $I4_1/amd$ symmetry (the structure of
Cs-IV) are both metallic~\cite{Cmca-4_structure,Pickard,McMahon_2011_structure_searching}.
DFT with the PBE functional predicts that $Cmca$-12 is stable up to 385
GPa, $Cmca$-4 is stable in the range 385--490 GPa, and $I4_1/amd$ is
stable from $490$ GPa up to pressures beyond 1 TPa.

DFT studies of high-pressure phases of hydrogen have been performed
using several approximate density functionals
\cite{Pickard_phase_IV,Chen,Morales_2013,me2,Cohen_2013} and a
significant dependence of the results on the functional has been
noted.  The enthalpy differences between phases are so small that changes
of only a few meV per proton can make a noticeable difference to the
phase diagram.  It is therefore important to use an accurate approach
for calculating the energies of the competing phases.  We have chosen
to use the DMC method \cite{Ceperley_DMC,matthew} to calculate the
static lattice energies.  This method solves the many-electron
Schr\"{o}dinger equation, is in principle exact for the ground
states of the hydrogen atom and molecule, and rigorously excludes
self-interaction errors.  It is likely that DMC provides a
considerably more accurate description of the energetics of hydrogen
than the currently available exchange-correlation density functionals.

We used the \textsc{casino} code \cite{casino} to perform fixed-node
DMC simulations with a trial wave function of the Slater-Jastrow (SJ)
form,
\begin{equation}
  \Psi_{\rm SJ}({\bf R}) = \exp[J({\bf R})] \det[\psi_{n}({\bf r}_i^{\uparrow})] 
  \det[\psi_{n}({\bf r}_j^{\downarrow})],
\label{eq1}
\end{equation}
where ${\bf R}$ is a $3N$-dimensional vector of the positions of the $N$
electrons, ${\bf r}_i^{\uparrow}$ is the position of the $i$'th spin-up
electron, ${\bf r}_j^{\downarrow}$ is the position of the $j$'th
spin-down electron, $\exp[J({\bf R})]$ is a Jastrow factor, and
$\det[\psi_{n}({\bf r}_i^{\uparrow})]$ and $\det[\psi_{n}({\bf
  r}_j^{\downarrow})]$ are Slater determinants of spin-up and spin-down
one-electron orbitals.  These orbitals were obtained from DFT
calculations performed with the plane-wave-based \textsc{quantum
  espresso} code \cite{QS}, employing a norm-conserving pseudopotential
constructed within DFT using the local density approximation (LDA)
exchange-correlation functional.  The choice of exchange-correlation
functional used to generate the orbitals has almost no effect on the DMC
energies of solid hydrogen phases \cite{me3,LDA-PBE0}.  Earlier work also
suggests that using a pseudopotential has only a small impact on results
for high-pressure solid hydrogen
\cite{McMahon_2011_structure_searching}.  We chose a very large
basis-set energy cut-off of 300 Ry to approach the complete basis set
limit \cite{sam}, as detailed in the Supplemental Material~\cite{sup}.
The plane-wave orbitals were transformed into a localized ``blip''
polynomial basis \cite{blip}.  Our Jastrow factor consists of polynomial
one-body electron-nucleus and two-body electron-electron terms, the
parameters of which were optimized by minimizing the variance of the
local energy at the variational Monte Carlo (VMC) level
\cite{varmin1,varmin2}.  The QMC calculations were performed with
simulation cells containing $N=128$ protons.  We used twist-averaged
boundary conditions with 24 randomly chosen twists to reduce the
single-particle finite-size effects \cite{twistav}.  We have corrected
our results for the effects of using finite simulation cells, employing
the approach described in Refs.~\cite{Chiesa} and \cite{Neil}.  The
residual finite-size effects are estimated to lead to errors in the
enthalpy differences between phases of less than 5 meV per proton. The
finite-size corrections are detailed in the Supplemental
Material~\cite{sup}. The statistical errors in our data are smaller than
the reported accuracy~\cite{sup}.

\begin{figure}[ht]
    \centering
    \includegraphics[width=0.4\textwidth]{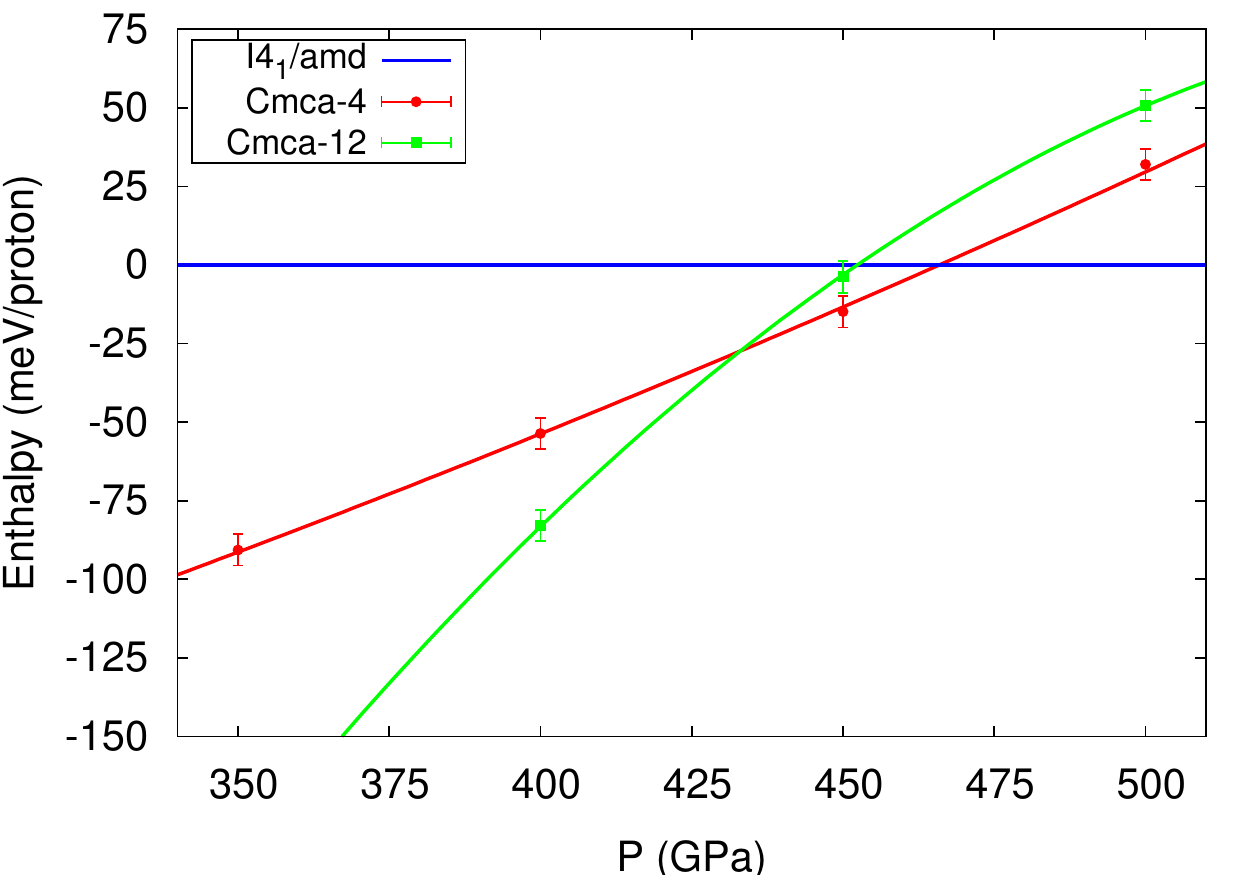} \\
    \includegraphics[width=0.4\textwidth]{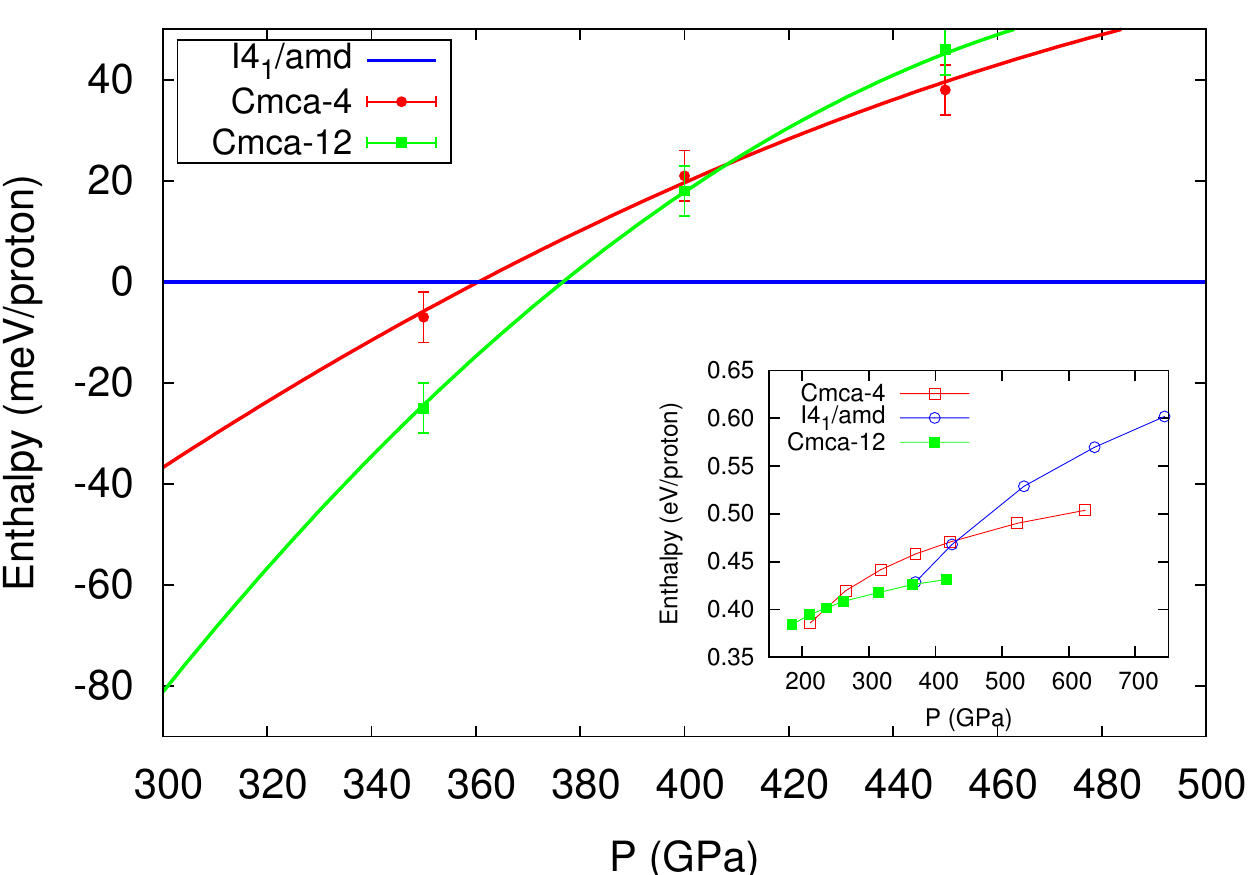} \\
    \caption{ \label{static}(color online).  Enthalpy as a function of
      pressure for the $Cmca$-12, $Cmca$-4, and $I4_1/amd$ phases,
      relative to the $I4_1/amd$ phase. \textit{Upper}: static phase diagram from DMC
      calculations.  \textit{Lower}: phase diagram including the ZP
      enthalpy from the harmonic and anharmonic vibrational
      calculations.  The inset shows the
      harmonic ZP enthalpy as a function of pressure.}
\end{figure}

The enthalpy was evaluated by fitting a polynomial to the
finite-size-corrected DMC energy as a function of volume and
differentiating the fit. The resulting enthalpy-pressure relations are
shown in the upper plot of Fig.~\ref{static}.  At the static lattice
level we find a transition from $Cmca$-12 to $Cmca$-4 at 431 GPa and a
transition from the molecular $Cmca$-4 to atomic $I4_1/amd$ structure
at about 465 GPa. DMC calculations predict that the $Cmca$-$4$ phase 
is significantly less stable than in PBE DFT.

To explore the accuracy of the DMC results further we performed
calculations with trial wave functions incorporating an inhomogeneous
backflow (BF) transformation \cite{Lopez_Rios_backflow}, which modifies
the nodal surface of the wave function and can introduce additional
correlation effects.  The BF wave functions give energies lower than the
SJ wave functions by about 19--13 meV per proton with increasing density
for $Cmca$-$4$, while the $I4_1/amd$ energies were lowered by about 17
meV per proton, approximately independently of density.  The energy
reduction in the molecular $Cmca$-$4$ phase is slightly larger than in
the $I4_1/amd$ atomic phase, but the energy reductions of the $Cmca$-4
and $I4_1/amd$ phases are almost the same in the region of the phase
transition.  We conclude that the introduction of BF correlations does
not significantly alter our results. Further details of the effects of
BF are described in the Supplemental Material~\cite{sup}.

The above results are based on static lattice calculations
in which the vibrational motion of the protons has been
neglected.  We have also performed calculations of the ZP enthalpy arising
from the proton motion using (a) the quasiharmonic approximation and
(b) a VSCF approach that enables the calculation of anharmonic
vibrational energies \cite{Tomeu1}.  The quasiharmonic phonon
calculations were performed with the PBE functional using both the
supercell finite displacement method and 
density-functional perturbation theory (DFPT) as implemented in
\textsc{quantum espresso} \cite{QS}.

Previous calculations of quasiharmonic proton ZP energies in solid
hydrogen have encountered significant numbers of unstable phonon modes
\cite{Pickard,McMahon_2011_structure_searching} at high pressures.  We
found that the $Cmca$-$12$ and $I4_1/amd$ structures had stable modes
at the supercell sizes considered.  For $Cmca$-$4$ we found a small
unstable region around the $\Gamma$ point which was further reduced,
but not entirely eliminated, by using cells with up to 256
protons. This small unstable region does not affect our
estimates of the ZP energy, as shown in the Supplemental
Material~\cite{sup}. As illustrated in the inset of the lower panel of Fig.~\ref{static},
the proton ZP enthalpy of all three phases increases with pressure.

Systems of light atoms with weak bonding often exhibit large
vibrational amplitudes, which are likely to give rise to anharmonic
vibrations.  There is evidence for the importance of anharmonicity in
hydrogen, especially in the high-density regime \cite{Morales_2013,
  McMahon_review_RMP}.
Utilizing our recently developed variational VSCF
scheme \cite{Tomeu1,helium}, we have calculated the anharmonicity of the
proton ZP motion of both the molecular and atomic phases. We use the
principal axes approximation to map the Born-Oppenheimer energy
surface along independent but anharmonic vibrational modes
\cite{Tomeu1, Jung}, and solve the resulting equations within a VSCF
scheme \cite{Tomeu1, Bowman}. We also calculate the contribution from
phonon-phonon two body coupling in the most anharmonic modes to estimate
the effects of these terms on the anharmonic vibrational energy \cite{sup}.
The Born-Oppenheimer energy surface is mapped within plane-wave DFT
using the \textsc{castep} code \cite{castep}. 
By comparing the energies of the highest- and
lowest-energy modes with those of the static lattice,
we estimate that our choice of computational parameters leads to energy
differences between frozen phonon configurations that are converged to
within $10^{-4}$ eV/proton.
 All calculations were performed with
the PBE functional and supercells containing $96$ and $108$ atoms.
Supercell-size convergence tests indicate that the anharmonic ZP
energy correction is accurate to within $1$ meV/proton
for all three phases. Further details of the VSCF calculations are
given in the Supplemental Material~\cite{sup}.  We performed
calculations for the atomic $I4_1/amd$ phase at pressures of $P=400$,
$500$, and $600$ GPa, obtaining anharmonic corrections of $-7.2$,
$-8.1$, and $-7.3$ meV/proton, respectively.  Similar calculations for
the molecular $Cmca$-$4$ phase at pressures of $P=400$ and $500$ GPa
give anharmonic corrections of $+8.7$ and $+8.3$ meV/proton,
respectively. Calculations at $P=400$ GPa for the $Cmca$-$12$ structure
lead to an anharmonic correction of $+4.0$ meV/proton.  The
anharmonic corrections to the proton ZP energy lower the energy of
the atomic phase and raise the energy of the molecular phases.

\begin{figure}
    \centering
    \includegraphics[width=0.4\textwidth]{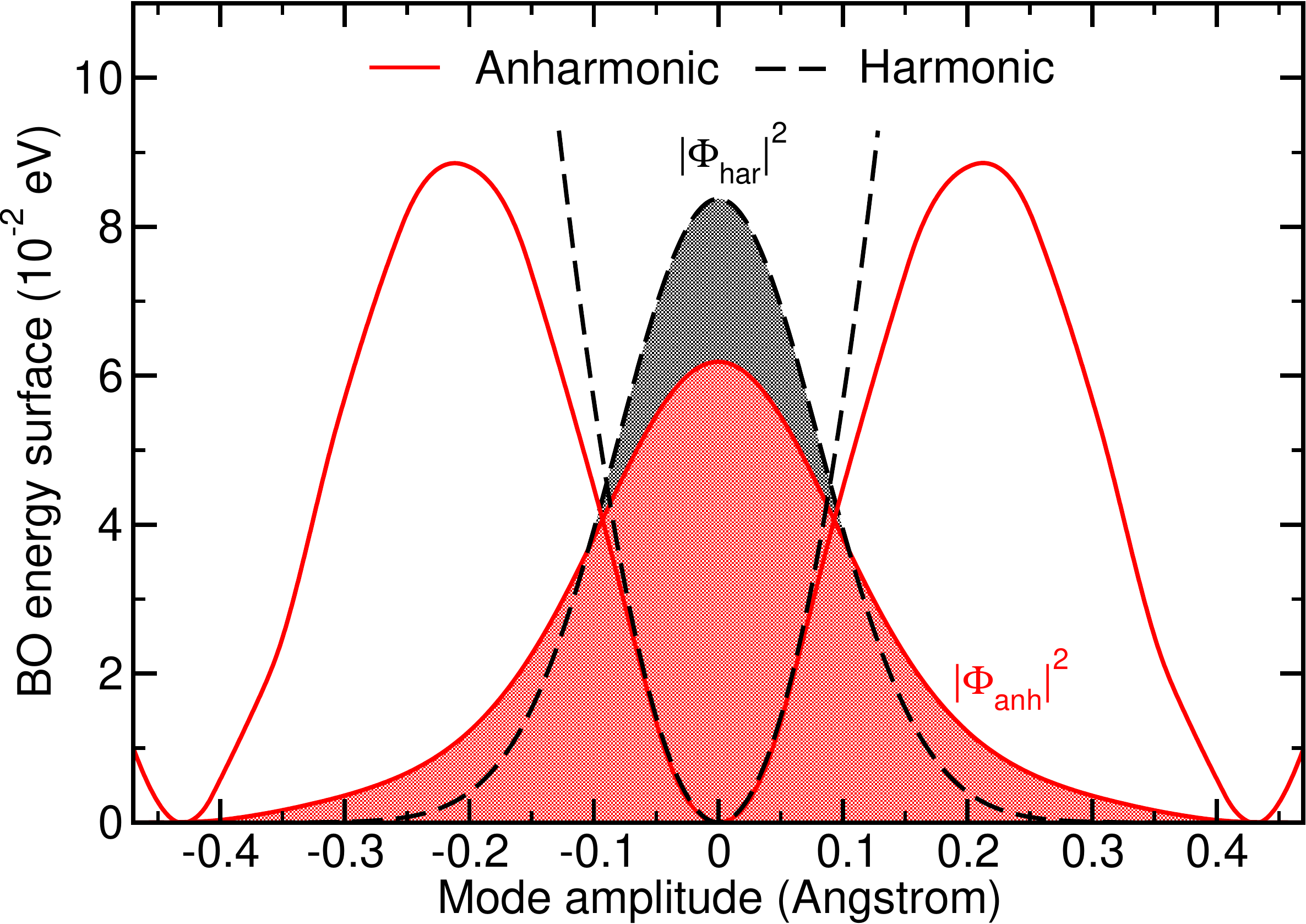} \\
    \hspace{0.1cm}
    \includegraphics[width=0.3\textwidth]{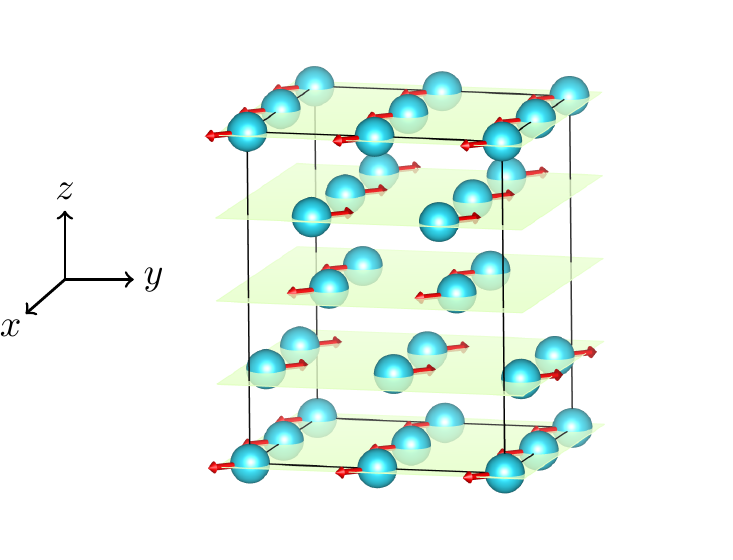} \\
    \caption{ \label{fig:anharmonic}(color online). \textit{Upper}: 
      Harmonic (black dashed) and anharmonic (red solid) Born-Oppenheimer energy
      surfaces and corresponding wave function densities 
      $|\Phi_{\mathrm{har}}|^2$ and $|\Phi_{\mathrm{anh}}|^2$
      for an optical mode at the $\Gamma$-point of the $I4_1/amd$
      structure. \textit{Lower}: A supercell of the $I4_1/amd$
      structure with arrows indicating the proton motion corresponding
      to the phonon mode in the upper figure. Alternate planes move 
      in anti-phase. }
\end{figure}

As an example of the vibrational properties of the $I4_1/amd$ structure, 
we plot in Fig.~\ref{fig:anharmonic} the Born-Oppenheimer energy surface and the
corresponding anharmonic wave function density at $P=500$~GPa for a
$\Gamma$-point optical phonon of the $I4_1/amd$ structure, and a
comparison with the harmonic quantities.  The $I4_1/amd$ structure can
be viewed as a sequence of four stacked planes with square lattices as
shown in the lower panel of Fig.~\ref{fig:anharmonic}.  The mode
corresponds to an in-plane motion of the protons, where alternate
stacked planes oscillate in opposite directions.  The minima of the
anharmonic potential are separated by about $0.427$~\AA\@ in real space. 
Adjacent minima correspond to
equivalent $I4_1/amd$ structures connected by this in-plane proton
motion.  This is the mode with the largest anharmonicity in the 
$I4_1/amd$ structure. 

We note that the fermionic nature of the protons has not been taken into
account at either the harmonic or anharmonic levels.  In order to
estimate the effects of this approximation, we consider the real space
amplitude of the motion of the protons about their equilibrium
positions.  The root-mean-square atomic amplitude in the $I4_1/amd$
phase at $P=500$~GPa is $\sqrt{\langle u^2\rangle}=0.099$~\AA\@, which
is much smaller than the nearest neighbour distance of $a=0.99$~\AA\@.
This indicates a small overlap between the proton wave functions and
justifies the neglect of their quantum statistics.  The
  Lindemann criterion \cite{lindemann} for the melting of a solid is
  usually taken to be $\sqrt{\langle u^2\rangle}/a\gtrsim0.1$, but
  for quantum melting a value of $\sqrt{\langle
    u^2\rangle}/a\gtrsim0.25$ is considered more accurate
  \cite{quantum_melting1,quantum_melting2,quantum_melting3}.
From our
data~\cite{sup}, $\sqrt{\langle u^2\rangle}/a=0.1$ at $T=0$~K, which
is not in the regime of a zero-temperature quantum
liquid~\cite{Ashcroft2}.    
The inclusion of quantum statistics would increase the kinetic energy
of the liquid and make it even less favorable at zero temperature. The
Lindemann criterion applied to the corresponding atomic phase for the
heavier deuterium would lead to a higher melting temperature.  The
Lindemann criterion cannot definitely answer the question of whether the
ground-state atomic phase is a metallic solid or a quantum liquid, but
it suggests that the melting temperature is higher than zero. These
conclusions should be compared with recent PIMD results at $P=500$~GPa
\cite{Chen}, which suggest a melting temperature of $160$~K.

We estimate the dynamical enthalpy as the sum of the static DMC
enthalpy and the ZP enthalpy calculated using the
quasiharmonic approximation corrected by the VSCF scheme to account
for the effects of anharmonicity. 
The contribution to the total pressure from the ZP motion for the
atomic $I4_1/amd$ phase increases with pressure from $25$ to $45$ GPa
over the pressure range of the inset in the lower panel of Fig.~\ref{static},
while the ZP pressure of the molecular $Cmca$-4 and $Cmca$-12 phases
increases from $19$ to $24$ GPa and $10$ to $13.5$ GPa, respectively. 
As shown in the dynamical phase diagram of Fig.~\ref{static}, the transition from the $Cmca$-12 to
the atomic $I4_1/amd$ phase occurs at $374$ GPa, and there is no 
stability region for the $Cmca$-$4$ phase. 

At the static level, the phase diagrams of hydrogen and deuterium are 
identical. At the dynamic level and using the quasiharmonic 
approximation, the deuterium ZPE is calculated by dividing the hydrogen 
ZPE by $\sqrt{2}$. As illustrated in the Supplemental Material~\cite{sup},
the dynamical phase diagram of deuterium shows that the molecular-to-atomic phase 
transition happens at a pressure of 390 GPa, also through a structural
transformation from molecular $Cmca$-$12$ to atomic $I4_1/amd$. Therefore,
the molecular-to-atomic phase transition is fairly isotope-independent.

We find that our DMC results are essentially independent of the
exchange-correlation (XC) functional used to calculate the orbitals
for the trial wave function.  However, the value of the proton ZP
energy depends on the choice of XC functional.  To investigate the
effect of this we have recalculated the harmonic ZP enthalpy for the
$Cmca$-4, $Cmca$-12 and $I4_1/amd$ structures using the BLYP XC
functional~\cite{blyp}, as detailed in the Supplemental
Material~\cite{sup}, which gives significantly different results from
the PBE functional at the static lattice level. The phase diagram
including the effects of the ZP harmonic enthalpy calculated with the
BLYP functional leads to a reduction of $28$~GPa in the transition
pressure for hydrogen dissociation, compared to the PBE-based phase
diagram. The differences in dissociation pressure due to the flavour
of XC functional used for the treatment of atomic vibrations do not
affect the qualitative results presented in this work, and only have a
limited quantitative effect.

In conclusion, we have studied the dissociation of solid molecular
hydrogen at the static lattice and dynamical lattice levels.
At the static lattice level our calculations give a transition from the
$Cmca$-12 molecular phase to the $Cmca$-4 molecular phase at
$P=431$~GPa, and a transition to the $I4_1/amd$ atomic phase at $465$
GPa.
At the dynamical level the molecular $Cmca$-12 phase
transforms directly to the atomic $I4_1/amd$ phase at 374 GPa. 
The limited precision of our calculations prevents us from stating
categorically that the $Cmca$-$4$ phase does not exist, but the 
pressure range over which it might exist is very narrow.
The atomization pressure is close to being within range of DAC experiments~\cite{dubrovinsky_dac}.
Therefore the low temperature molecular-to-atomic phase transition
of high pressure hydrogen might be observable experimentally.  
By comparing the dynamical phase diagrams of hydrogen and deuterium, 
we predict that the molecular-to-atomic phase transition is
almost isotope-independent.
The proton ZP vibrational energies increase with pressure
and the anharmonic contribution leads to an increase in the
vibrational energy of the molecular $Cmca$-$4$ and $Cmca$-$12$ phases and a decrease in that of
the $I4_1/amd$ atomic phase.
Our results suggest that quantum melting of hydrogen would occur at finite temperature.
Since metallic hydrogen is thought to be present in large amounts in the
interiors of Jupiter, Saturn, and some extra-solar planets, planetary 
models should consider incorporating our
prediction of the existence of an atomic metallic state at lower
pressures than previously assumed.

We acknowledge the financial support of the UK Engineering and Physical Sciences Research Council 
under grants EP/I030190/1 and EP/I030360/1, the use of the HECToR computing
facilities (the UK national supercomputing service), the Imperial College 
London High Performance Computing Centre,
and the Cambridge High Performance Computing Service. We acknowledge support
from the Thomas Young Centre under grant TYC-101


\begin{thebibliography}{99}
\bibitem{LDA-PBE0} The Cmca-12 phase is metallic in the LDA but
  insulating up to 373 GPa in $GW$ calculations. To check the accuracy
  of our metallic LDA-based trial functions for Cmca-12, we also used trial
  functions from hybrid PBE0 calculations with a much larger band
  gap. The difference between the two DMC energies was less than 2 meV
  per proton at all pressures considered.
\end{thebibliography}
\end{document}